\documentclass[11pt,twoside]{jltp} %jltp
\usepackage{overcite}
\usepackage{psfig}
\title{Having It Both Ways: Distinguishable Yet Phase-Coherent Mixtures
of Bose-Einstein Condensates}

\author{E.~A. Cornell, D.~S. Hall, M.~R. Matthews, and C.~E. Wieman}

\address{JILA, National Institute of Standards and Technology\\ and
  Department of Physics, University of Colorado, Boulder, Colorado}

\runninghead{E.~A. Cornell {\it et al.}}{Distinguishable Yet Phase-Coherent 
Condensates}
       
\begin{document}

\maketitle

\begin{abstract}
  We have begun a series of experiments on mixed bosonic quantum
  fluids. Our system is mixed Bose-Einstein condensates in dilute
  Rb-87. By simultaneously trapping the atoms in two different
  hyperfine states, we are able to study the dynamics of component
  separation and of the relative quantum phase of two interpenetrating
  condensates. Population can be converted from one state to the other
  at a rate that is sensitive to the relative quantum phase.

\end{abstract}

%Include this space if you do not use sections in your document.
%\vspace{0.3in}

\section{INTRODUCTION}
\label{introduction}

The observation of Bose-Einstein condensation (BEC) in a dilute gas
\cite{jila,mit,rice} has led to a new generation of experiments in
quantum fluids.  Historically, in the study of quantum fluids, some of
the most intriguing behavior has been found in the behavior of fluid
mixtures.  We have thus been motivated to pursue a series of
mixed-condensate experiments in Rubidium-87.  \cite{myatt, mike,
  david1, david2} The experimental tools and the theoretical
world view of atomic physics are almost entirely disjoint from those
of traditional low-temperature physics, and this lack of overlap has
led to some confusion. The purpose of this paper is to review some of
our early results on mixed condensates, and to provide a qualitative
exegesis of the theoretical and experimental techniques that underpin
our work.  Of necessity we have omitted here much of the detail
provided in the original papers. \cite{myatt,mike,david1,david2} The
MIT BEC group also has begun experiments in mixed condensates.
\cite{mitwork}

The experiments described here all begin with a sample of
approximately $5 \times 10^5$ spin-aligned, Bose-condensed Rb-87 atoms
at a temperature of less than 50~nK, confined in a 3-d, harmonic
magnetic potential at density $1 \times 10^{14}/{\mathrm {cm}}^3$.
The number density of noncondensed atoms is a factor of 30 or more
smaller.  In this paper we will skip the details of refrigeration ---
the technique is a hybrid method that combines many of the notable
advances \cite{cool1980} in atom cooling from the 1980s: laser cooling
and trapping, magnetic trapping, and evaporative cooling.  The
synthesis of these various approaches was begun at JILA from 1989--1992
\cite {vaporcell, lowroad}, but it was only after considerable
additional development \cite{darkspot,cmot,mitevap,top} that, in 1995,
the parts worked well enough together to permit the first observation
of BEC in dilute atomic gases \cite{jila,mit,rice}.  In the current
version of our apparatus, the cooling process is automated and
reliable, much like a (well-behaved) dilution refrigerator. A freshly
prepared condensate is available for study about once per minute.  The
isolation of the sample from the 300~K environment (less than 1~cm
away) is excellent --- thermal equilibration times with the environment
are on the order of centuries --- but residual heating and the
collisional decay of the condensate limit the duration of an
experiment on any given condensate to a few seconds.

\begin{figure}
\centerline{\psfig{file=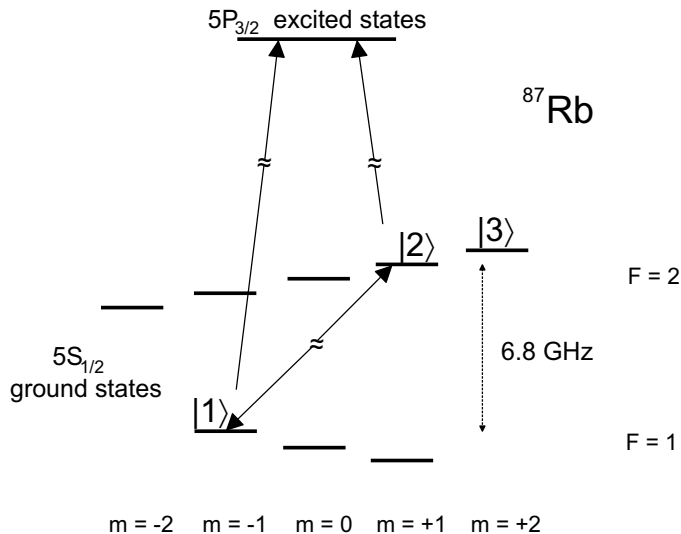,height=2.75in}}
%
%\framebox[5in]{\rule[1.125in]{0in}{1.125in}}
%\makebox[5in]{\rule[1.125in]{0in}{1.125in}}
\caption{The relevant hyperfine and Zeeman structure of Rb-87 in a 
magnetic field. The atoms are moved coherently from state $\left|1\right>$  to
state $\left|2\right>$  with a microwave pulse\protect\cite{twophoton}. 
Imaging is accomplished via 
absorption in optical transitions to the excited states. }  
\label{fig:levels}
\end{figure}

The ground state of an Rb-87 atom is split in a magnetic field into
eight levels by hyperfine and Zeeman interactions
(Fig.~\ref{fig:levels}). The states labelled $\left|1\right>$,
$\left|2\right>$, and $\left|3\right>$ can be trapped at a local
minimum in the magnetic field; it is with states $\left|1\right>$ and
$\left|2\right>$ ($\left|F=1,m=-1\right>$, and $\left|F=2,
  m=1\right>$, respectively) that we have performed most of our
mixed-fluids experiments. \cite {decaysup,opttrap} The other relevant
internal structure of the Rb-87 atom is a set of excited-state levels
1.6~eV above the ground state (Fig.~\ref{fig:levels}).  We probe the
spatial distribution of the atoms by imaging the absorption on an
optical transition to these levels.  The natural 6~MHz linewidth of
the optical transition is much less than the energy splitting between,
for instance, states $\left|1\right>$ and $\left|2\right>$. With the
correct choice of laser frequency, we can image the spatial
distribution of either state $\left|1\right>$ or of state
$\left|2\right>$, or, if we prefer, of the combined density of the two
states.\cite{mike}

The paradoxical aspect of the Rb-87 mixed-condensate system is that it
combines two traits that are usually thought of as mutually
exclusive: (i) the $\left|1\right>$ and $\left|2\right>$ components of
the condensate are distinguishable, in the usual senses of the term
--- the components do not at experimental energy scales spontaneously
interconvert, and the components can be selectively detected and
imaged; and (ii) the components possess a measurable relative quantum
phase, which evolves at a rate proportional to the difference in
chemical potential between the two fluids, and which can, under the
right circumstances, be read out, much as the current across a
Josephson junction reads out the relative quantum phase across it.
Interconversion (and phase read-out) {\it can} be driven via a
stimulated process and therefore can be turned on and off. The system
thus bears considerable resemblance to the idealized pair of
condensates linked by a removable Josephson junction that Leggett
\cite{leggett} and others have envisioned in various thought
experiments.

In Section~2 we will discuss distinguishable-fluid experiments; in
Section~3 we discuss simple phase-coherent experiments; in Section~4
we discuss experiments that combine the two aspects; and in Section~5
we mention various future experiments with weak coupling.

\section{INTERPENETRATING, DISTINGUISHABLE FLUIDS}
\label{interpenetrating}

Species $\left|1\right>$ and $\left|2\right>$ meet both usual criteria
for calling fluids ``distinguishable.'' One can in the literal sense
distinguish one density distribution from the other, simply by
adjusting the frequency of the probe laser so that one species or the
other casts its shadow onto our imaging CCD array
(Fig.~\ref{fig:levels}). This measurement, as performed in
Refs.~\citen{mike,david1,david2}, is destructive, but as has been shown
\cite{MITdisp} it needn't be so. In any case the fact that we can
prepare condensates with a high repetition rate and a large degree of
reproducibility means that the destructive aspect of the imaging is
not very important to us: we observe the time-evolution of the clouds
simply by repeating the experiment many times with increasing
dwell-times.  The second sense of ``distinguishable'' is that particles
do not spontaneously interconvert.  This requirement is enforced in
our system by the enormous difference in internal energies between the
two states. In the absence of applied electromagnetic fields, the
6.8~GHz hyperfine energy (Fig.~\ref{fig:levels}) is a million times
larger than any other energy scale of the condensate system. The
energy released by a {\it single} atom converting from state
$\left|2\right>$ to state $\left|1\right>$, if that energy were to be
distributed thermally through our sample, is sufficient to drive
the {\it entire sample} out of the Bose-condensed state.  The fact
that we don't observe our condensates melting during our measurements
is a guarantee that spontaneous interconversion is not a factor in the
system.

The positions of the two fluids are determined by the their respective
confining magnetic potentials.  The magnetic moments of the two states
are nominally the same, but due to various small effects --- gravity,
the nuclear magnetic moment, the onset of nonlinearity in the Zeeman
shifts, and subtle dynamical effects \cite{top,spie} of our magnetic
trap --- the location of the two minima of the confining potentials
$V_1(r)$ and $V_2(r)$ can be adjusted to be either exactly coincident
or slightly offset.  For the work described here, the confining
potentials are axially symmetric and harmonic, with single-particle
oscillation frequencies of $\omega_z/2\pi = 60$~Hz, and $\omega_r/2\pi
= 21$~Hz.  With the spatial offset between the two potentials set to
zero, and in the absence of interspecies interactions, the two fluids would
be completely interpenetrating.  In fact, inter- (and intra-) species
interactions are a dominant factor in determining the density
distributions. The dynamics are well-described in the mean-field
language of coupled Gross-Pitaevskii equations for the order
parameters: \cite{twobectheory}

\begin{equation}
  i\hbar\dot{\Phi}_1 = \left(-\frac{\hbar^2}{2m}\nabla^2 + V_1 + u_1
    |\Phi_1|^2 + u_{12}|\Phi_2|^2\right)\Phi_1
\label{gpe1a}
\end{equation}
and
\begin{equation}
  i\hbar\dot{\Phi}_2 = \left(-\frac{\hbar^2}{2m}\nabla^2 + V_2 +
    V_{\mathrm{hf}} + u_2|\Phi_2|^2 + u_{21}|\Phi_1|^2 \right)\Phi_2
\label{gpe1b}
\end{equation}
where $m$ is the mass of the Rb atom, $V_{\mathrm{hf}}$ is the
magnetic field-dependent hyperfine splitting between the two states in
the absence of interactions, $u_i = 4\pi\hbar^2a_i/m$ and $u_{ij} =
4\pi\hbar^2a_{ij}/m$, $|\Phi_i|^2$ is the condensate density, and the
intraspecies and interspecies scattering lengths are $a_i$ and
$a_{ij}=a_{ji}$. Note that these equations do not allow for species
interconversion; they conserve independently the number of atoms in
each species.

Considerable theoretical work has already gone into studying
ground-state solutions \cite{twobectheory} and small-amplitude
excitations \cite{twobecex} of the order parameters. We review here
the results most relevant to the current experiments. As the total
number of atoms trapped increases, the quantum kinetic energy term
$\nabla^2 \Phi$ becomes less significant, until it is almost
negligible for the numbers used in double-condensate experiments.
Most theoretical treatments neglect this term, thus making the
so-called Thomas-Fermi approximation. In this case, the steady-state,
single-component condensate density profile in a parabolic potential
will have an inverted parabaloid density profile, with density
tapering smoothly from the peak at cloud center to zero at its edge.
Due to an accidental degeneracy \cite{myatt,greenestuff} in the Rb-87
scattering system, $a_1 \approx a_{12} \approx a_2$.  For this special
case, one can see by inspection of Eqs.~\ref{gpe1a} and~\ref{gpe1b}
that the steady-state total density of a two-component BEC will also
have an inverted parabola form, even though the relative densities of
the components may have intricate structure.  It turns out there is a
critical value \cite{twobectheory} for the interaction term $a_{12}^c
= \sqrt{a_1 a_2} $.  For $a_{12}>a_{12}^c$, two components should have
little spatial overlap --- they should spontaneously separate in the
trap, while for $a_{12}<a_{12}^c$, there should be a relatively large
region of interpenetration.  In Rb-87, the scattering lengths are
known at the 1\% level to be in the proportion
$a_1:a_{12}:a_2::1.03:1:0.97$, with the average of the three being
55(3)~\AA~\cite{mike,greenestuff,Burke_PVT}.

Since the mutual interaction term is within 1\% of its critical value,
anything that breaks the symmetry between the two species becomes
important.  For instance, the fact that $a_1 > a_2$ means it is
energetically favorable for the $\left|1\right>$ atoms to
preferentially move towards the lower density region at the periphery
of the cloud, forming a spherical shell around
$\left|2\right>$,\cite{twobectheory} but since the difference between
$a_1$ and $a_2$ is small, even a rather minor vertical offset in the
spatial centers of $V_1$ and $V_2$ will result in the components'
separating up-and-down, rather than radially in-and-out.  In
steady-state, then, one expects the clouds to be largely, although not
completely \cite{kinsepeng} spatially separated.  Further, because the
steady-state energetics only modestly favor component separation, one
would expect that in the ensuing dynamical behavior, small
oscillations about the steady-state configuration\cite{twobecex} will
feature a number of ``soft'' modes, with frequencies low compared to
the excitations of the total density.

\begin{figure}
\centerline{\psfig{file=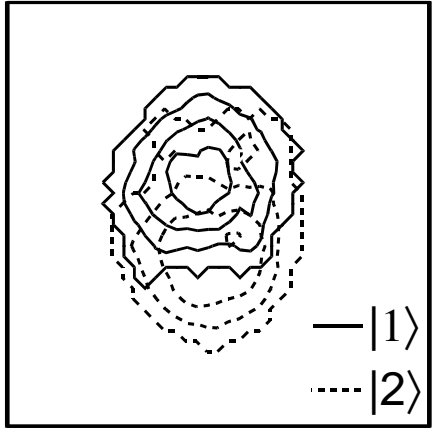,height=2.25in}}
%
%\framebox[5in]{\rule[1.125in]{0in}{1.125in}}
%\makebox[5in]{\rule[1.125in]{0in}{1.125in}}
\caption{Contour plot of the expanded \protect\cite{expand}  
  density distribution after the two components have undergone
  separation and their relative motions have damped.  The image is 182
  microns square.  Before expansion the size of the total distribution
  was 40 by 14 microns. Note that a considerable region of
  interpenetration remains.  }
\label{fig:densityprofile}
\end{figure}

All of these qualitative theoretical expectations have been borne out
in preliminary experiments.\cite{david1}.
Figure~\ref{fig:densityprofile} shows the density profile
\cite{expand} of two condensates after they have come to steady-state.
In this measurement, the potentials have been offset a distance equal
to only 3\% of the overall sample size, but the resulting component
separation is much larger.  Note however that there does remain a
region of overlap, which is useful for work described below.  The
$\left|2\right>$ component is originally created (as described in
Section~3.2, below) such that it completely overlaps the $\left|1\right>$
component, and we have observed\cite{david1} the time evolution of the
component separation as it relaxes to the steady-state, separated
condition. During this time, the overall density profile is
essentially unperturbed.  The period of the resulting damped
oscillations of the relative positions of the two components is about
30~ms. In comparison, the period of the lowest-order nontrivial
excitation of the total density profile is about 25~ms.

If we do not deliberately break symmetry of the confining potentials
in the experiment but rather keep them as perfectly overlapped as
possible, we see, at early times, the onset of the expected
inward-outward component separation.  At longer times the inner ball
of $\left|2\right>$ atoms does not stay well-centered in the hollow
shell of $\left|1\right>$ atoms, but drifts horizontally towards a
preferred side. There appears to be an as-yet uncharacterized
imperfection in the trap that breaks the radial symmetry of the
experimental environment.

The Rubidium-87 mixture is well-suited to a variety of experiments in
mixtures of dilute bose gases.  There are a host of experimental
parameters to explore: the radial:axial aspect ratio of the confining
potential (from 3:1 to greater than 1:30, depending on the form of
magnetic trap used); the relative number in the two components; and the
temperature. The critical temperature of a particular component
depends on the number of atoms in that state.  Thus, one can arrange
for one component to remain in the condensed state while the other one
converts from the normal to the condensed state or \emph{vice-versa}.
All of these changes will have profound effects on the steady-state
configuration and the spectrum of collective excitations. For example,
imagine an experiment to study the effects of finite temperature on
the excitations.\cite{finiteT} At finite temperature, there will be in
effect four different co-trapped fluids: normal and condensate
fractions of both components. The normal fraction of a given component
will find the condensate fraction of its own component twice as
repulsive as it finds the condensate fraction of a dissimilar
component (due to exchange terms.)  Thus if one chooses a potential in
which condensate $\left|1\right>$ and condensate $\left|2\right>$
tend to spatially separate, one can have a situation in which the
normal fraction of component $\left|1\right>$ tends to preferentially
collect at the spatial location of condensate $\left|2\right>$, and
{\it vice-versa}. What will the collective excitation spectrum look
like? If we were to describe these excitations with the word
``sound,'' what ordinal number should modify it?

\section{PHASE COHERENCE}
\label{phasecoherence}

In the absence of any applied coupling the 6.8~GHz energy separation
between the two states is enormous, but the presence of a oscillating
magnetic field tuned close to the hyperfine splitting changes the
situation dramatically.  An atom can readily be transferred from one
state to the other --- the difference in energy is absorbed from (or
stimulated into) the coherent field of 6.8~GHz photons.
\cite{twophoton} That the population transfer resulting from the
application of a coupling field is sensitive to the relative phase of
the two states is well-known in microwave spectroscopy \cite{ramsey}
and NMR and is easy to understand in the case of a single, two-level
atom.
 
\subsection{Single-Atom Case}
\label{singleatom}

The Schr\"odinger equation \cite{dshplease} for the internal state of
a two-level atom (or of any two-level quantum system) in the presence
of a coupling drive at frequency $\omega_{\mathrm{rf}}$ is

\begin{equation}
  i\dot{A}_1= \omega_1A_1 + \frac{\Omega}{2} e^{i \omega_{\mathrm{rf}}t}A_2
\label{schr1a}
\end{equation}

\begin{equation}
  i\dot{A}_2= \omega_2A_2 + \frac{\Omega^*}{2} e^{-i
  \omega_{\mathrm{rf}}t}A_1
\label{schr1b}
\end{equation}

\noindent{where $A_i$ is the probability amplitudes to be in state 
  $\left|i\right>$, $\hbar\omega_i$ is the internal energy of state
  $\left|i\right>$, and $\Omega$ is the amplitude of the coupling.  We
  assume the detuning $\delta \equiv
  \omega_{\mathrm{rf}}-\omega_2+\omega_1$ is small, such that
  $|\delta|\ll \omega_{\mathrm{rf}}$, and concentrate on the case
  that the coupling drive is off ($\Omega=0$) most of the time and
  on only for brief pulses of duration $\tau$, such that $1/\tau \gg
  \left|\delta\right|$.  When we discuss ``phase'' in this context, we
  are talking about the real quantities $\alpha_1$, $\alpha_2$, or
  $\alpha_{\mathrm{rf}}$ defined in the relations $A_1 = |A_1|e^{i \alpha_1}$,
  $A_2 = |A_2|e^{i \alpha_2}$, and $\Omega e^{- \omega_{\mathrm{rf}}t}
  = |\Omega |e^{i \alpha_{\mathrm{rf}}}$.
  
  There is one particularly useful special-case solution of
  Eqs.~\ref{schr1a} and~\ref{schr1b}.  A pulse with duration and
  amplitude such that $|\Omega|\tau= \frac{\pi}{2}$ is known as a
  ``$\frac{\pi}{2}$-pulse.'' If the atom is initially in state
  $\left|1\right>$ with unit probability ({\it i.e.}, $|A_1|=1$ and
  $|A_2|=0$), then after the application of the pulse the atom will
  have equal probability of being in either state ({\it i.e.},
  $|A_1|=1/\sqrt{2}$ and $|A_2|=1/\sqrt{2}$).  The
  $\frac{\pi}{2}$-pulse thus creates a 50-50 coherent superposition of
  the two states, also ``writing'' a particular initial relative
  phase. After the pulse is applied, the relative phase evolves at the
  rate $\omega_2-\omega_1$.  The accumulated phase can be measured by
  a second $\frac{\pi}{2}$-pulse.\cite{ramsey} When a
  $\frac{\pi}{2}$-pulse is applied to a two-level system in 50-50
  coherent superposition, the population transferred by the pulse from
  state $\left|1\right>$ to state $\left|2\right>$ is proportional to
  $\cos(\alpha_1-\alpha_2-\alpha_{\mathrm{rf}})$, where the phases
  $\alpha$ refer to the total phase accumulated during the dwell time
  $T$ between the two pulses.  One measures the transition probability
  from state 1 to state 2 by repeating the experiment for different
  $T$; the result is ``Ramsey fringes,'' a cosine at the detuning
  frequency $\delta$. This technique is known as the method of
  separated oscillatory fields \cite{ramsey} or ``twin-pulse''
  spectroscopy.
        
\subsection{Application to Order Parameters}
\label{orderparam}

\begin{figure}
\centerline{\psfig{file=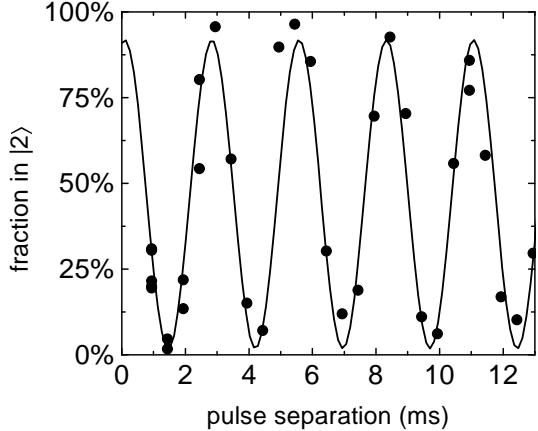,height=2.25in}}
%
%\framebox[5in]{\rule[1.125in]{0in}{1.125in}}
%\makebox[5in]{\rule[1.125in]{0in}{1.125in}}
\caption{Population transfer from $\left|1\right>$  to  $\left|2\right>$ 
resulting from twin $\frac{\pi}{2}$-pulse 
coupling pulses, as a function of delay between the two pulses. 
The first pulse prepares the condensate
in an equal superposition of the two states, the second pulse induces 
further population transfer sensitive to the relative phase that has
evolved during the pulse separation time. The coupling drive is detuned
from the energy difference between the two condensates by about 360~Hz.}  
\label{fig:ramseyfringes}
\end{figure}

Now it becomes clear how we modify Eqs.~\ref{gpe1a} and~\ref{gpe1b} to
include a coupling field applied to a bose condensate:

\begin{equation}
  i\hbar\dot{\Phi}_1 = 
  \left(-\frac{\hbar^2}{2m}\nabla^2 + V_1 + 
                u_1 |\Phi_1|^2 + u_{12}|\Phi_2|^2\right)\Phi_1
  + \frac{\hbar\Omega(t)}{2}e^{i\omega_{\mathrm{rf}}t}\Phi_2
\label{gpe2a}
\end{equation}
and
\begin{equation}
  i\hbar\dot{\Phi}_2 = \left(-\frac{\hbar^2}{2m}\nabla^2 + 
                V_2 +
    V_{\mathrm{hf}} + u_2|\Phi_2|^2 + u_{21}|\Phi_1|^2 \right)\Phi_2 +
  \frac{\hbar\Omega(t)}{2}e^{-i\omega_{\mathrm{rf}}t}\Phi_1
\label{gpe2b}
\end{equation}

\noindent{The probability amplitudes $A_i$ of Eqs.~\ref{schr1a}
  and~\ref{schr1b} have now become field amplitudes $\Phi_i$, and the
  physics has become correspondingly richer. One can still perform
  experiments, however, that explore simple limits.}

In one such experiment, we begin with atoms in the pure
$\left|1\right>$, relaxed to a steady-state, near-pure condensate with
spatial distribution described by $\Phi_{0}$.  Next we apply a
$\frac{\pi}{2}$-pulse, transferring half the population to state
$\left|2\right>$.  Immediately after the $\frac{\pi}{2}$-pulse, the
two-state system can be described by $\Phi_1 =
\Phi_{0}e^{i\mu_1t}/\sqrt {2}$, and $\Phi_2 =
\Phi_{0}e^{i\mu_2t}/\sqrt {2}$, where $\mu_1$ and $\mu_2$, which
differ by about the hyperfine frequency, are the chemical potentials
of the two components. We know however that this perfectly overlapping
state is not the steady-state of the mixed BEC system; this is in fact
the point of departure for the relaxation-to-equilibrium experiments
described in Section 2.1. Within a matter of a few tens of
milliseconds the components will separate, but on shorter time scales,
the condensates don't have time to realize that they are a pair of
mutually repulsive quantum fluids. One could equally well describe the
system at short times as a single condensate containing about a
million atoms, each in an internal, coherent superposition governed by
Eqs.~\ref{schr1a} and~\ref{schr1b}. Figure~\ref{fig:ramseyfringes}
shows the result of applying a second $\frac{\pi}{2}$-pulse
immediately, and then measuring the final number of atoms in the
$\left|2\right>$ state.  The high-contrast Ramsey fringes that result
are much as one would expect from a similar experiment performed on,
for instance, an atomic beam.\cite{ramsey} This ``conventional''
response of condensates at short times to coupling drives has also
been seen in condensate Rabi oscillations.\cite{mike,outputcoupler}

\section{COHERENT FLUIDS}
\label{coherentfluids}

\begin{figure}
\centerline{\psfig{file=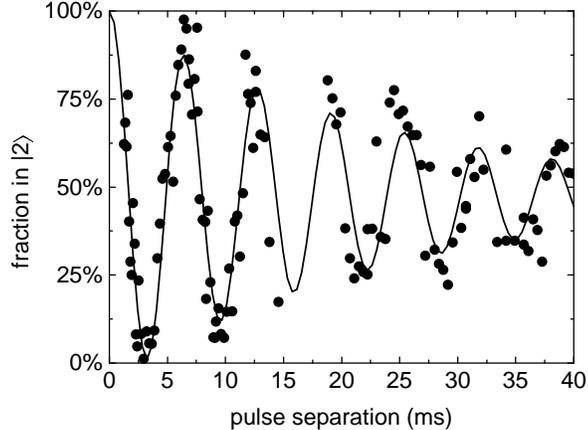,height=2.25in}}
%
%\framebox[5in]{\rule[1.125in]{0in}{1.125in}}
%\makebox[5in]{\rule[1.125in]{0in}{1.125in}}
\caption{Population transfer from $\left|1\right>$  to $\left|2\right>$  
  resulting from twin $\frac{\pi}{2}$-pulse coupling pulses, as a
  function of delay between the two pulses.  Results are similar to
  those shown in Fig.~\protect\ref{fig:ramseyfringes}, except that
  over the longer duration the components physically separate --- the
  loss of spatial overlap reduces the contrast ratio of the Ramsey
  fringes. }
\label{fig:ramseyfading}
\end{figure}

If the system is allowed to evolve longer after the first
$\frac{\pi}{2}$-pulse, so that the components begin to separate, it is
no longer very useful to describe the system as a single condensate of
atoms in a superposition state. One should think of them instead as
distinct fluids evolving their separate ways, albeit (as we shall see)
with a well-defined relative phase.  The effects of component
separation show up clearly in twin-pulse spectroscopy:
Figure~\ref{fig:ramseyfading} shows some condensate Ramsey fringes
collected over longer periods. The loss of contrast in the fringes is
not due to inhomogenous broadening of the atomic transition, at least
not in the conventional sense.  Rather, it arises from the dwindling
spatial overlap between the two components, which in turn arises from
the mutual repulsion of the components, as we saw in Section~2. During
component separation the components are of course moving with respect
to each other, and therefore there is a gradient in the relative phase
across the sample; it is no longer so simple to define a single
quantity as the phase differerence between the two clouds.

\begin{figure}
\centerline{\psfig{file=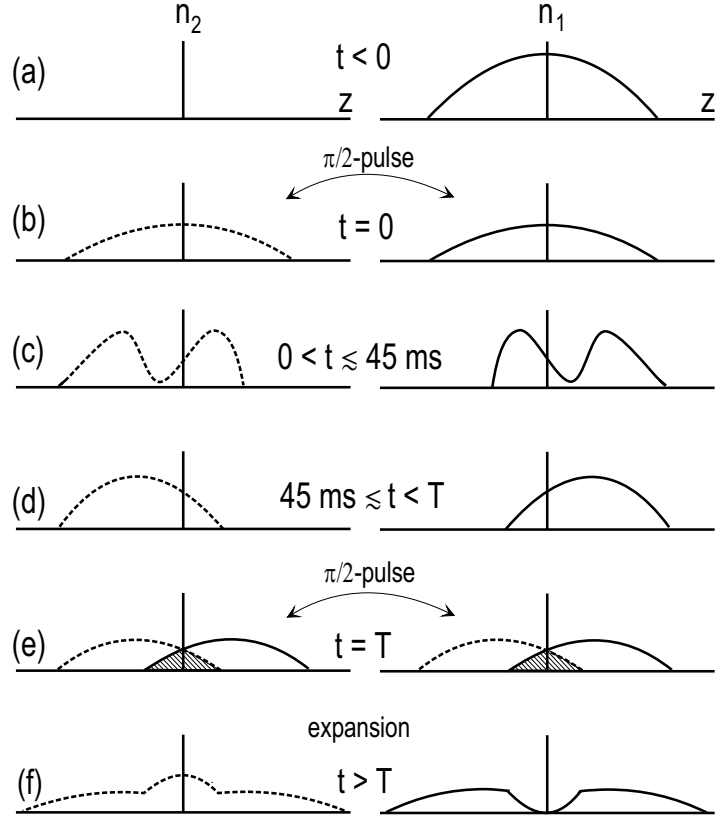,height=4.25in}}
%
%\framebox[5in]{\rule[1.125in]{0in}{1.125in}}
%\makebox[5in]{\rule[1.125in]{0in}{1.125in}}
\caption{A schematic \protect\cite{david2} of the condensate interferometer. (a) The
  experiment begins with all of the atoms in condensate
  $\left|1\right>$ at steady-state. (b) After the first
  $\frac{\pi}{2}$-pulse, the condensate has been split into two
  components with a well-defined initial relative phase. (c) The
  components begin to separate in a complicated fashion due to mutual
  repulsion as well as a $0.4~\mu$m vertical offset in the confining
  potentials (see also Fig.~3 of Ref.~\protect\citen{david1}). (d) The
  relative motion between the components eventually damps with the
  clouds mutually offset but with some residual overlap. Relative
  phase continues to accumulate between the condensates until (e) at
  time $T$ a second $\frac{\pi}{2}$-pulse remixes the components; the
  two possible paths by which the condensate can arrive in one of the
  two states in the hatched regions interfere.  (f) The cloud is
  released immediately after the second pulse and allowed to expand
  for imaging. In the case shown, the relative phase between the two
  states at the time of the second pulse was such as to lead to
  destructive interference in the $\left|1\right>$ state and a
  corresponding constructive interference in the $\left|2\right>$
  state.}
\label{fig:interscheme}
\end{figure}

At still longer inter-pulse times, the results of the twin-pulse
experiments become once again explainable in terms of a single
relative phase.  After the component-separation process has gone to
completion, the relative motion of the condensates damps, and
presumably the gradients of the relative phase vanish as well. As
shown in Fig.~\ref{fig:densityprofile} above, the post-separation
clouds preserve a reasonable amount of spatial overlap.  As shown
schematically in Fig.~\ref{fig:interscheme}, the application the
second $\frac{\pi}{2}$-pulse pulse at this time acts as sort of
recombiner on a separated-arm atom-interferometer\cite{pritchard}.
Depending on the relative phase between the two components, we can see
(Fig.~\ref{fig:postdampfringe}) destructive or constructive
interference in the region where the components overlap.  Thus, the
double-pulse experiment with a long-time between the pulses acts as a
condensate interferometer that allows us to determine the relative
phase between two largely separated condensate components.

\begin{figure}
\centerline{\psfig{file=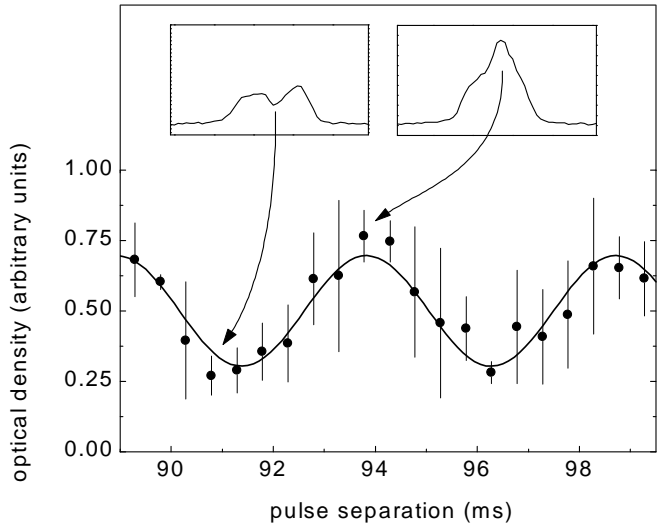,height=2.75in}}
%
%\framebox[5in]{\rule[1.125in]{0in}{1.125in}}
%\makebox[5in]{\rule[1.125in]{0in}{1.125in}}
\caption{The value of the condensate density in the
  $\left|2\right>$ state is extracted at the center of the overlap
  region (inset) and plotted as a function of $T$. Each point
  represents the average of 4~separate realizations and the thin bars
  denote the rms scatter in the measured interference for an
  individual realization.  The thick lines are sinusoidal fits to the
  data, from which we extract the 
  angular frequency $\mu_2-\mu_1-\omega_{\mathrm{rf}}$. }  
\label{fig:postdampfringe}
\end{figure}

We find the contrast ratio manifest in Fig.~\ref{fig:postdampfringe}
quite remarkable.  It is evidence that the two components have a
well-preserved memory of their relative phase \emph{even after} there
has been damping of the external degrees of freedom, with which the
internal states are entangled. The phase of the condensate evidently
has a robustness one would not expect in a single-particle experiment.
This is clearly an area worthy of extensive study -- how robust is the
relative phase, as a function of temperature, of relative equilibrium
separation, of time between pulses?  How well can the relative phase
survive a ``nondestructive'' observation \cite{MITdisp}, which can
take many sequential images of the same condensate? In single-atom
interferometry, the connection between the availability of
``which-path'' information and the loss of coherence has been
thoroughly examined.\cite{pritchard_dephase} How do these results
generalize to condensates?

\section{ADDITIONAL DIRECTIONS}
\subsection{Josephson Effects}
In the discussion so far the coupling drive was operated in the very
short-pulsed mode.  If instead it is applied as a weak, continuous
drive, it has correspondingly smaller bandwidth, such that it is
resonant in certain regions of the condensate, and off-resonant
elsewhere. Population transfer can be constrained to occur only within
a particular spatial region, for instance right at the overlap region,
and can show oscillatory but nonlinear behavior.  We are performing
experiments \cite{us_soon} in this regime, and are exploring the
close analogy \cite{jaime, josephson} between the resulting population
transfer and a nonlinear Josephson junction.

\subsection {Number States}
In experiments to date, the relative phase appears random if measured
more than 150~ms after the condensates are split. This could be due to
intrinsic decohering mechanisms, or it could be due to technical
instabilities.  It is diverting to contemplate, however, creating a
condensate pair for which from the very beginning the relative phase
is \emph{intrinsically} unpredictable.  The conjugate variable to the
relative phase is the number difference $\Delta N =N_1-N_2$.  For
experiments with relatively small total $N$, say less than 2000, it
may be possible to create states for which $N_1 \approx N_2 \approx
N/2$, with the uncertainty in $\Delta N$ less than plus or minus
1~atom.  Such a state, analogous to a highly squeezed state of light,
will have a completely undefined relative phase and could be useful
for performing sub-shotnoise spectroscopy, \cite{kasevich} and
illustrates some instructive paradoxes of quantum measurement theory.
\cite{juha}

\section*{ACKNOWLEDGMENTS}
This research is supported by the NSF, the ONR, and NIST.


\begin{thebibliography}{9}
\bibitem{twophoton} A transition from the $\left|1\right>$ state
  ($F=1, m=-1$) to the $\left|2\right>$ state ($F=2, m=1$) requires
  two units of angular momentum. The microwave coupling drive must
  therefore be composed of two photons. As described in
  Ref.~\protect\citen{mike}, we use a two-photon drive well-detuned from the
  intermediate $F=1,m=0$ state.  For all practical purposes, one may
  model the coupling drive as a single photon transition with a
  particular detuning and effective strength.
\bibitem{decaysup} The exception is Ref.~\protect\citen{myatt}, for
  which states $\left|1\right>$ and $\left|3\right>$ were used. In
  Ref.~\protect\citen{myatt} it was discovered that, due to an
  accidental degeneracy in scattering lengths, spin-exchange processes
  are suppressed in Rb-87 (see Ref.~\protect\citen{greenestuff}). In
  the absence of this degeneracy, one would expect condensate mixtures
  of different hypefine levels to have a very short lifetime.
\bibitem{opttrap}The MIT 
BEC group has confined condensates in optical potentials
(Ref.~\protect\citen{mitwork}) and shown that
experiments can be performed on mixtures  on the $m=1, m=0$ and $m=-1$
sublevels of the $F=1$ hyperfine state of sodium.
\bibitem{expand} 
In order to get better spatial
resolution of the clouds, we utilize a pre-imaging expansion technique. 
The confining potentials are discontinuously turned off, which allows the 
self-repulsion of the cloud to expand rapidly the density distribution.
We know from both numerical simulation and laboratory experience that
this process does little other than linearly expand the space the cloud
occupies. In particular, the relative positions of the two components
in the cloud are preserved.
\end{thebibliography}
\end{document}